\documentclass[twocolumn]{aastex63}

\usepackage{bm}
\usepackage{lipsum, babel}

\newcommand{\HI}{H{\sc~i}}

\newcommand{\s}{\text{s}}

\newcommand{\g}{\text{ g }}
\newcommand{\C}{\text{ C }}

\newcommand{\W}{\text { W }}

\newcommand{\K}{\text{K}}

\newcommand{\J}{\text{ J }}

\newcommand{\Mag}{\text{ mag }}

\newcommand{\pd}[2]{\frac{\partial #1}{\partial #2}}

\newcommand{\E}{\vec{E}}

\shorttitle{Late and Rapid Reionization}
\shortauthors{Cain et al.}

\graphicspath{{./}{figures/}}

\begin{document}

\title{A short mean free path at $z=6$ favors late and rapid reionization by faint galaxies}

\correspondingauthor{Christopher Cain}
\email{ccain002@ucr.edu}

\author[0000-0001-9420-7384]{Christopher Cain}
\affiliation{Department of Physics and Astronomy, University of California, Riverside, CA 92521, USA}

\author{Anson D'Aloisio}
\affiliation{Department of Physics and Astronomy, University of California, Riverside, CA 92521, USA}

\author{Nakul Gangolli}
\affiliation{Department of Physics and Astronomy, University of California, Riverside, CA 92521, USA}

\author{George D. Becker}
\affiliation{Department of Physics and Astronomy, University of California, Riverside, CA 92521, USA}

\begin{abstract}

Recent measurements of the ionizing photon mean free path ($\lambda_{912}^{\rm mfp}$) at $5 < z < 6$ suggest that the IGM was rapidly evolving at those times.  We use radiative transfer simulations to explore the implications for reionization, with a focus on the short value of $\lambda_{912}^{\rm mfp} = 3.57^{+3.09}_{-2.14}$ cMpc/$h$ at $z=6$.  We introduce a model for sub-resolution ionizing photon sinks based on radiative hydrodynamics simulations of small-scale IGM clumping.  We argue that the rapid evolution in $\lambda_{912}^{\rm mfp}$ at $z=5-6$, together with constraints on the metagalactic ionizing background, favors a late reionization process in which the neutral fraction evolved rapidly in the latter half. We also argue that the short $\lambda_{912}^{\rm mfp}(z=6)$ points to faint galaxies as the primary drivers of reionizaton. Our preferred model, with $\lambda_{912}^{\rm mfp}(z=6) = 6.5$ Mpc/$h$, has a midpoint of $z= 7.1$ and ends at $z= 5.1$.  It requires 3 ionizing photons per H atom to complete reionization and a LyC photon production efficiency of $\log(f^{\rm eff}_{\rm esc} \xi_{\rm ion}/[\mathrm{erg}^{-1} \mathrm{Hz}]) = 24.8$ at $z>6$.  Recovering $\lambda_{912}^{\rm mfp}(z=6)$ as low as the measured central value may require an increase in IGM clumpiness beyond predictions from simulations, with a commensurate increase in the photon budget. 
\end{abstract}

\bigskip


\section{Introduction} \label{sec:intro}

There are several lines of evidence that reionization ended around $z=6$, or perhaps later.   The large Ly$\alpha$ forest opacity fluctuations at $z \geq 5.5$ have been attributed to neutral islands below $z=6$ \citep{Becker2015, Bosman2018, Eiliers2018, Kulkarni2019, Keating2019, Nasir2020, Qin2021}.   Other independent constraints from high-$z$ Ly$\alpha$ emitter (LAE) surveys and quasar damping wing analyses  also suggest a significantly neutral IGM at $z\sim 7$ \citep{Davies2018, Wang2020, Kashikawa2006, Schenker2012, Pentericci2014, Ono2011}. The reionization models invoked to explain these observations are consistent with the low value of Cosmic Microwave Background optical depth reported by Planck, $\tau_{\rm CMB} = 0.054 \pm 0.007$ \citep{Planck2018}.  Given large uncertainties in these measurements, however, it is unclear whether reionization was rapid or more extended in duration \citep[e.g.][]{Finkelstein2019}.  In this letter, we argue that recent measurements of the mean free path at $5 < z \leq 6$ point to a reionization process that was both late and rapid.  

The new evidence considered here was reported by \citet[][B21]{Becker2021}.  They extended direct measurements of the mean free path (MFP; $\lambda_{912}^{\rm mfp}$) to $z=6$ and found a rather low value of $\lambda_{912}^{\rm mfp}(z=6) = 3.57^{+3.09}_{-2.14}$ cMpc/h.  This disfavors at the $97\%$ level even the low MFP predicted by the ``Low $\tau_{\rm CMB}$" model of \citet{Keating2020}, in which the IGM was still $20\%$ neutral at $z=6$ and reionization ended at $z\approx 5.3$.  The B21 result suggests that ionizing photon sinks played a larger role than previous models have captured, and/or that the IGM was even more neutral at $z=6$. 

Recent theoretical work has demonstrated that modeling the sinks is complicated by the interplay between self-shielding and the hydrodynamic response of the IGM to photoheating \citep{Park2016, Daloisio2020}.  Before reionization, the gas clumps down to its Jeans scale, which in the $\Lambda$CDM cosmology can be as low as $\sim1$ kpc. After ionization fronts (I-fronts) sweep through, the local density structure ``relaxes" by Jeans smoothing and photoevaporative processes, evolving to a less clumpy state over a timscale $\Delta t \sim 200$ Myr.  State-of-the-art radiative hydrodynamics (RHD) simulations have yet to fully bridge the scale gap between the sinks and the $\gtrsim 100$ Mpc boxes necessary to converge on reionization observables \citep{Iliev2014}.  In addition, the largest dynamic ranges have been achieved with moment-based radiative transfer (RT) methods, which may be numerically inaccurate in the sinks \citep{Wu2021}. 

A key addition of this work is that we have developed a new sub-grid model for the sinks based on the study of \citet[][D20]{Daloisio2020}.  We have incorporated this into a new ray tracing RT code, giving our reionization simulations a formal dynamic range of over 5 orders of magnitude in scale.  In this letter, we present first results from this new simulation framework and we use them to interpret the B21 measurements. 

This letter is organized as follows.  In \S~\ref{sec:numerical}, we describe our numerical methods.  In \S~\ref{sec:results} we present our results and we conclude in \S~\ref{sec:conc}. 

\section{Numerical Methodology}
\label{sec:numerical}

We ran RT simulations of reionization in a $(200h^{-1}~\mathrm{cMpc})^3$ volume using our new ray tracing code, which we will describe in detail elsewhere (Cain \& D'Aloisio in prep.).  We use a coarse-grained uniform RT grid with $N_{\rm}=200^3$.  The premise is to use a pre-run suite of highly resolved (small volume) RHD simulations to model the opacity evolution in each $(1h^{-1}~\mathrm{Mpc})^3$ RT cell during reionization.   
 
 \subsection{Coarse-grained RT}
Our monochromatic RT algorithm traces rays from source cells using an adaptive splitting and merging scheme based on \citet{Abel2002} and similar to the procedure of \citet{Trac2007}.  We track 48 directions (HEALPix level 1) and use the full speed of light because the commonly adopted reduced-speed approximation leads to inaccuracies near the end of reionization.  We propagate sub-grid ``moving-screen" I-fronts across the coarse cells with speed $v_{\rm IF} = \frac{F}{(1+\chi) n_{\rm H}}$, where $F$ is the incident ionizing flux, $n_{\rm H}$ is the proper H density, and $\chi = n_{\rm He}/n_{\rm H} \approx 0.082$ accounts for singly ionized helium.  We assume that ray $j$ intersecting cell $i$ contributes photoionizations over a path length $x^i_{\rm ion} \Delta s ^{ij}$, where $x^i_{\rm ion}$ is the cell ionized fraction and $\Delta s^{ij}$ is the total path length of the ray through the cell. We further assume that the MFP in the ionized part of the cell takes a locally uniform value, $\overline{\lambda}^i$.  The mean H photoionization rate there is 

\begin{equation}
    \label{eq:gammah1}
    \Gamma^i_{\rm HI} = \sum^{N_{\rm rays}}_{j=1}\frac{N^{ij}_{\gamma,0}  [1 - \exp(-x^i_{\rm ion} \Delta s^{ij}/\overline{\lambda}^i)]~/~\Delta t}{x^i_{\rm ion} V_{\rm cell}~/~\overline{\sigma}_{\rm HI} \overline{\lambda}^i },
\end{equation}
where $N_{\gamma,0}^{ij}$ is the number of photons incident on the cell in a time $\Delta t$ from ray $j$, $V_{\rm cell}$ is the cell volume, and $\overline{\sigma}_{\rm HI}$ is the H photoionization cross-section.   The over-bars denote frequency averages. To compute these quantities, we assume a specific intensity of $J_{\nu} \propto \nu^{-1.5}$ between 1 and 4 Ry, consistent with models of young metal-poor stellar populations \citep[e.g.][]{DAloisio2019}.  The numerator of eq. \ref{eq:gammah1} counts H ionizations per unit time.  The denominator gives the number of \HI\ atoms in ionized gas since $(\overline{\sigma}_{\rm HI} \overline{\lambda}^i)^{-1}$ is the $\Gamma_{\rm HI}$-weighted mean $n_{\rm HI}$. We tested this moving-screen framework against simulations of plane-parallel I-fronts similar to those in D20 (but with one domain) and found good agreement in the photon budget.  For comparison against Ly$\alpha$ forest measurements, we track temperatures on the RT grid using the approximate method of \citet{DAloisio2019}.  We adopt their fit for the post I-front temperature, $T_{\rm reion}$, and the subsequent thermal evolution is modeled using their Eq. 6.  

\begin{figure*}
    \centering
    \includegraphics[scale=0.14]{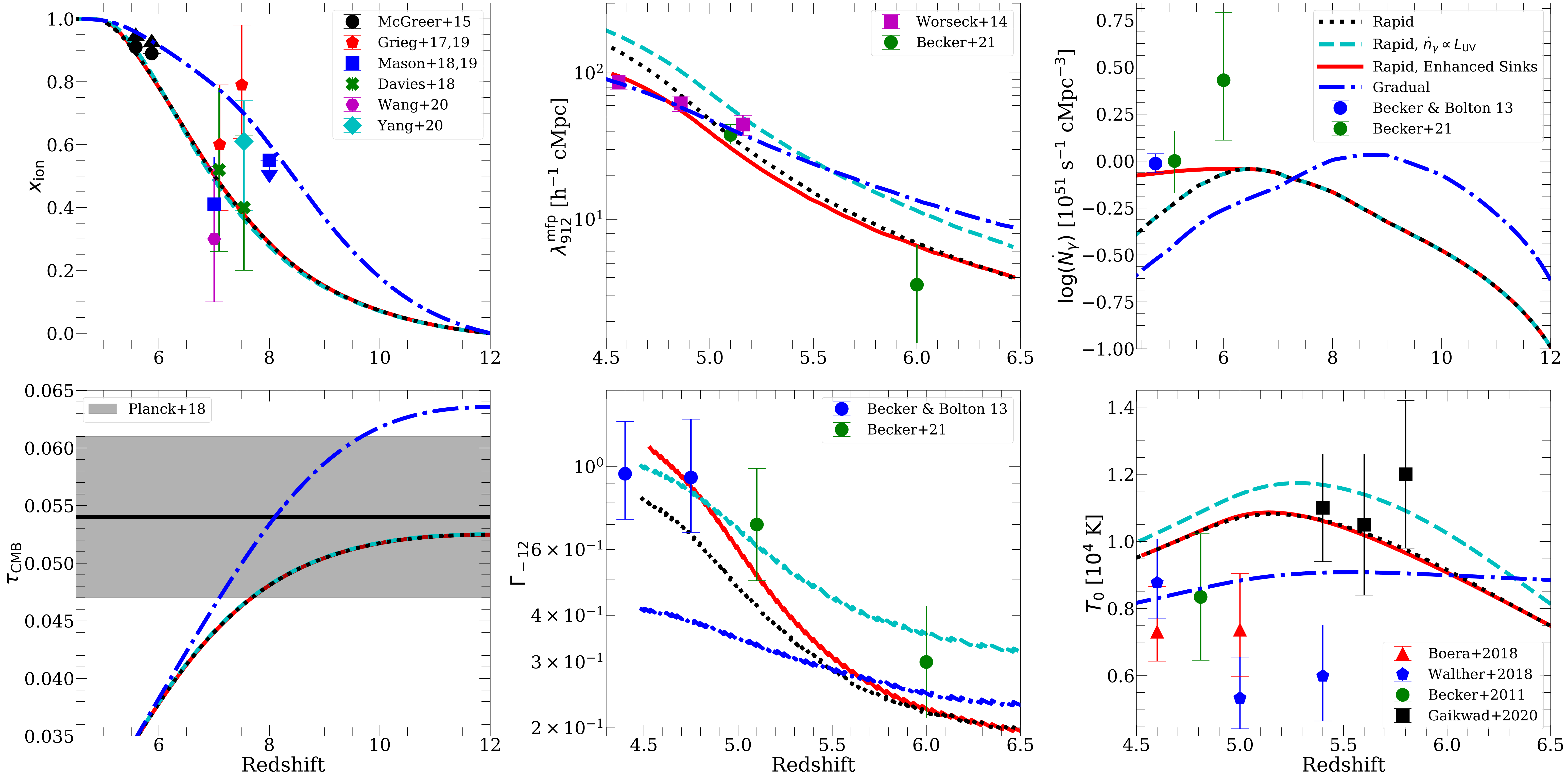}
    \caption{Reionization observables in our models. In clockwise order, starting from the top left, we show the ionized fraction, comoving MFP at 1 Ry, source comoving emissivity, IGM temperature at the cosmic mean density, H photoionization rate, and CMB optical depth.  We compare against a selection of recent observational constraints with 1$\sigma$ error bars.  } \citep{McGreer2015, Grieg2016,Grieg2019,Davies2018,Mason2018, Mason2019, Worseck2014, Planck2018, Becker2013, Becker2011,Walther2019,Boera2019,Gaikwad2020,Wang2020,Yang2020a}     
    \label{fig:ion_history}
\end{figure*}

\subsection{Sub-grid model for $\overline{\lambda}$}

We extract $\overline{\lambda}$ from the RHD simulations of D20, which use $N_{\rm dm} = N_{\rm gas} = N_{\rm rt} = 1024^3$ dark matter particles and gas/RT cells in a $(1h^{-1}~\mathrm{Mpc})^3$ box.  These simulations are parameterized by three environmental quantities: the reionization redshift, $z_{\rm re}$, the impinging $\Gamma_{\rm HI}$, which  quantifies the strength of the external ionizing background, and the box-scale linear over-density over its standard deviation, $\delta/\sigma$.  We have expanded the D20 suite to include all combinations of $\Gamma_{-12} \equiv \Gamma_{\rm HI}/10^{-12} {\rm s}^{-1} \in \{0.03, 0.3, 3.0\}$, $z_{\rm re} \in \{12, 8, 6\}$, and $\delta/\sigma \in \{-\sqrt{3}, 0, \sqrt{3}\}$. For cells reionized below $z_{\rm re} = 6$, we extrapolate logarithmically in cosmic time. We tested our extrapolation against a simulation with $z_{\rm re} = 5.2$ and found good agreement.  The frequency-averaged MFP is obtained using $\overline{\lambda^{-1}} = \langle n_{\rm HI} \Gamma_{\rm HI} \rangle_V / F$, where $\langle \ldots \rangle_V$ is a volume average. 

The D20 simulations track the self-shielding and hydrodynamic response of the IGM in the wake of I-fronts sourced by a steady background (constant $\Gamma_{\rm HI}$).  However, $\Gamma_{\rm HI}$ can evolve considerably in realistic environments. Using D20-style simulations with time-varying $\Gamma_{\rm HI}$, we have developed an empirical model for the evolution of $\overline{\lambda}$,
\begin{equation}
	    \label{eq:lambdamaster}
	    \frac{d \overline{\lambda}}{dt} = \pd{\overline{\lambda}}{t}\Big|_{\Gamma_{\rm HI}} + \pd{\overline{\lambda}}{\Gamma_{\rm HI}} \Big|_{t}\frac{d\Gamma_{\rm HI}}{dt} - \frac{\overline{\lambda} - \overline{\lambda}_{\rm eq}}{t_{\rm relax}}.
\end{equation}
The first term on the right is the quantity measured from the D20 simulations -- the evolution of $\overline{\lambda}$ at fixed $\Gamma_{\rm HI}$.  The second term captures the {instantaneous} change in $\overline{\lambda}$ with $\Gamma_{\rm HI}$.  The last term implements a relaxation timescale $t_{\rm relax}$ over which $\overline{\lambda}$ evolves towards an equilibrium value $\overline{\lambda}_{\rm eq}$  in response to a sudden increase in $\Gamma_{\rm HI}$.  The $\pd{\overline{\lambda}}{t}\Big|_{\Gamma_{\rm HI}}$ term and $\overline{\lambda}_{\rm eq}$ are interpolated from the expanded D20 suite.  When interpolating over $z_{\rm re}$, we correct for the I-front crossing time of the cell by averaging the opacity in the least and most recently ionized gas.  For $ \pd{\overline{\lambda}}{\Gamma_{\rm HI}} \Big|_{t} $, we assume a power law $\overline{\lambda} \propto \Gamma_{\rm HI}^\xi$, where $\xi = 2/3$.  This form is motivated by \citet{McQuinn2011}, the constraints of B21, as well as our D20-style calibration simulations.  Eq.~\ref{eq:lambdamaster} is integrated in fully ionized cells with $\Gamma_{-12} \geq 0.03$.  In partially ionized cells or those with $\Gamma_{-12} < 0.03$, we set $\overline{\lambda} = \overline{\lambda}_{\rm eq}$.  
Since $\overline{\lambda}$ depends on $\Gamma_{-12}$, we iterate Eqs.~\ref{eq:gammah1}-\ref{eq:lambdamaster} until convergence. The D20-style simulations that we used to calibrate Eq. \ref{eq:lambdamaster} included up to $\times 100$ impulsive increases in $\Gamma_{\rm -12}$ as well as more realistic cases with gradual evolution. Eq. (\ref{eq:lambdamaster}) captures $\overline{\lambda}(t)$ in the gradually time-varying-$\Gamma$ simulations to better than a few percent, while a straight interpolation over-estimates it by 10-15\%.  Our impulsive tests yielded similar levels of improvement.  Consistent with the results of these tests, we use $t_{\rm relax} = 100$ Myr.  

\subsection{Density fields and source models}

The RT sims were run on a coarse-grained version of the $(200 h^{-1}~\mathrm{Mpc})^3$ hydrodynamics simulation in \citet{DAloisio2018}, which employed a modified version of the code of \citet{Trac2004}.  Density fields and halo catalogs were saved at time intervals of 10 Myr from $z = 12 - 4.5$.  The halo mass functions are converged down to $2 \times 10^{10}$ h$^{-1}$M$_{\odot}$, more massive than the smallest halos believed to have contributed to reionization.  We extended our sources by generating sub-resolution halos down to $M_{\rm min}=10^9$ h$^{-1}$M$_{\odot}$ using the nonlinear biasing method of \citet{Ahn2015}, applied with the halo mass function of \citet{Trac2015}. (As discused below, we have also run tests with $M_{\rm min}=10^8$ h$^{-1}$M$_{\odot}$.)  We used two models for the ionizing photon production rate of each halo: $\dot{n}_{\gamma}=$ constant (dependent only on $z$) and $\dot{n}_{\gamma}\propto$ UV luminosity ($L_{\rm UV}$), where $\dot{n}_{\gamma}$ is the number of photons per unit time produced by a halo.  The former is our fiducial model, which assigns more weight to low-mass galaxies in the ionizing photon budget. As we will see, the motivation for this choice is that the short value of $\lambda^{\rm mfp}_{912}$(z=6) favors models in which reionization is driven by faint, less-biased sources. For the $L_{\rm UV}$ model, we abundance matched to the UV luminosity function of \citet{Finkelstein2019}. In both cases we chose the overall normalization of $\dot{n}_{\gamma}$ at each redshift to set the global emissivity history, $\dot{N}_{\rm \gamma}(z)$.  We varied $\dot{N}_{\rm \gamma}(z)$ by trial and error to find reionization models consistent with the Planck $\tau_{\rm CMB}$ constraints and the mean free path measurements.

\section{Results}
\label{sec:results}

\begin{figure}
    \centering
    \includegraphics[scale=0.19]{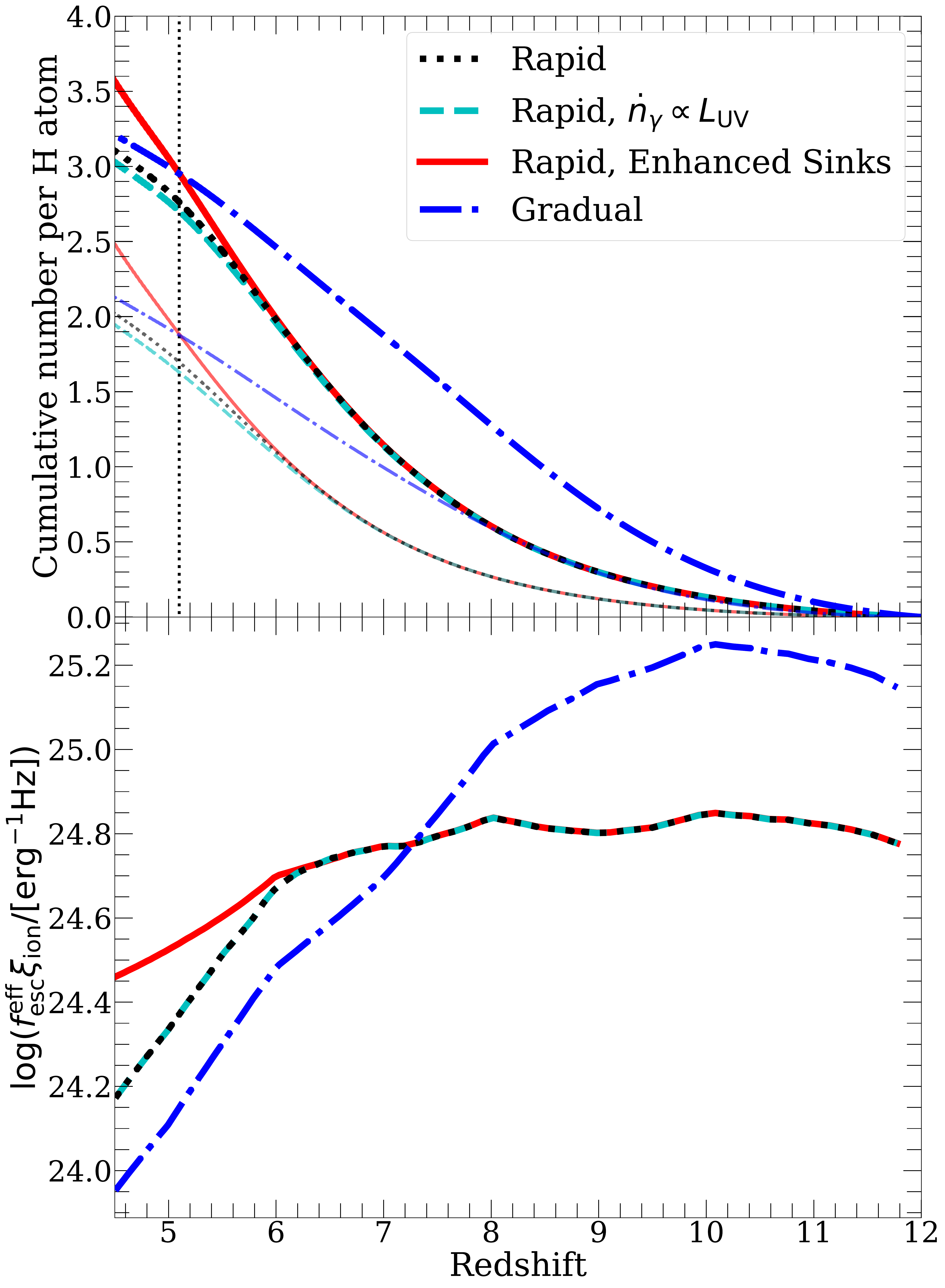}
    \caption{Top: Cumulative number of absorptions (thick curves) and recombinations (thin faded curves) per H atom for our models. The vertical line corresponds to the end of reionization.  Bottom: Product of the effective LyC escape fraction and ionizing efficiency for each of our models.  The rapid models are broadly consistent with observational constraints on $\xi_{\rm ion}$ provided that $f^{\rm eff}_{\rm esc} = 10-40 \%$ (see text). }
    \label{fig:budget}
\end{figure}

 In Fig. \ref{fig:ion_history} we show 4 models chosen to illustrate our conclusions. In clock-wise order, starting at the top left, the panels show the volume-weighted mean ionized fraction, comoving MFP at 1 Ry ($\lambda_{912}^{\rm mfp}$), comoving ionizing emissivity ($\dot{N}_{\gamma}$), temperature at the cosmic mean density ($T_0$), volume-weighted mean $\Gamma_{-12}$ in ionized gas, and the cumulative CMB optical depth.  We compare against an assortment of existing constraints. To calculate $\lambda_{912}^{\rm mfp}$, we traced 50,000 sight lines from random locations, created mock quasar absorption spectra, and fit to the model of \citet{Prochaska2009} and \citet{Worseck2014}.\footnote{Our sight lines do not start on QSOs, thus are not affected by biases from the proximity effect \citep{DAloisio2018, Becker2021}. When we anchored sight lines on massive halos, we found that the model of \citet{Prochaska2009} did not provide a good fit at rest-frame $900 \AA \lesssim \lambda < 912$\AA, owing to the local clustering of sources and the back-reaction of $\Gamma_{\rm HI}$ on $\lambda_{912}^{\rm mfp}$.  Excluding these wavelengths in the fit gave the un-biased $\lambda_{912}^{\rm mfp}$.  In this work, we assume that the observations represent the un-biased $\lambda_{912}^{\rm mfp}$. For reference, the biased $\lambda_{912}^{\rm mfp}(z = 6)$ is 9.3, 15.8, 15.1, 9.1 Mpc/h in the rapid, gradual, $\dot{n}_{\rm \gamma} \propto L_{\rm UV}$,  and enhanced sinks models, respectively.  } 

\begin{figure*}[t]
    \centering
    \includegraphics[scale=0.60]{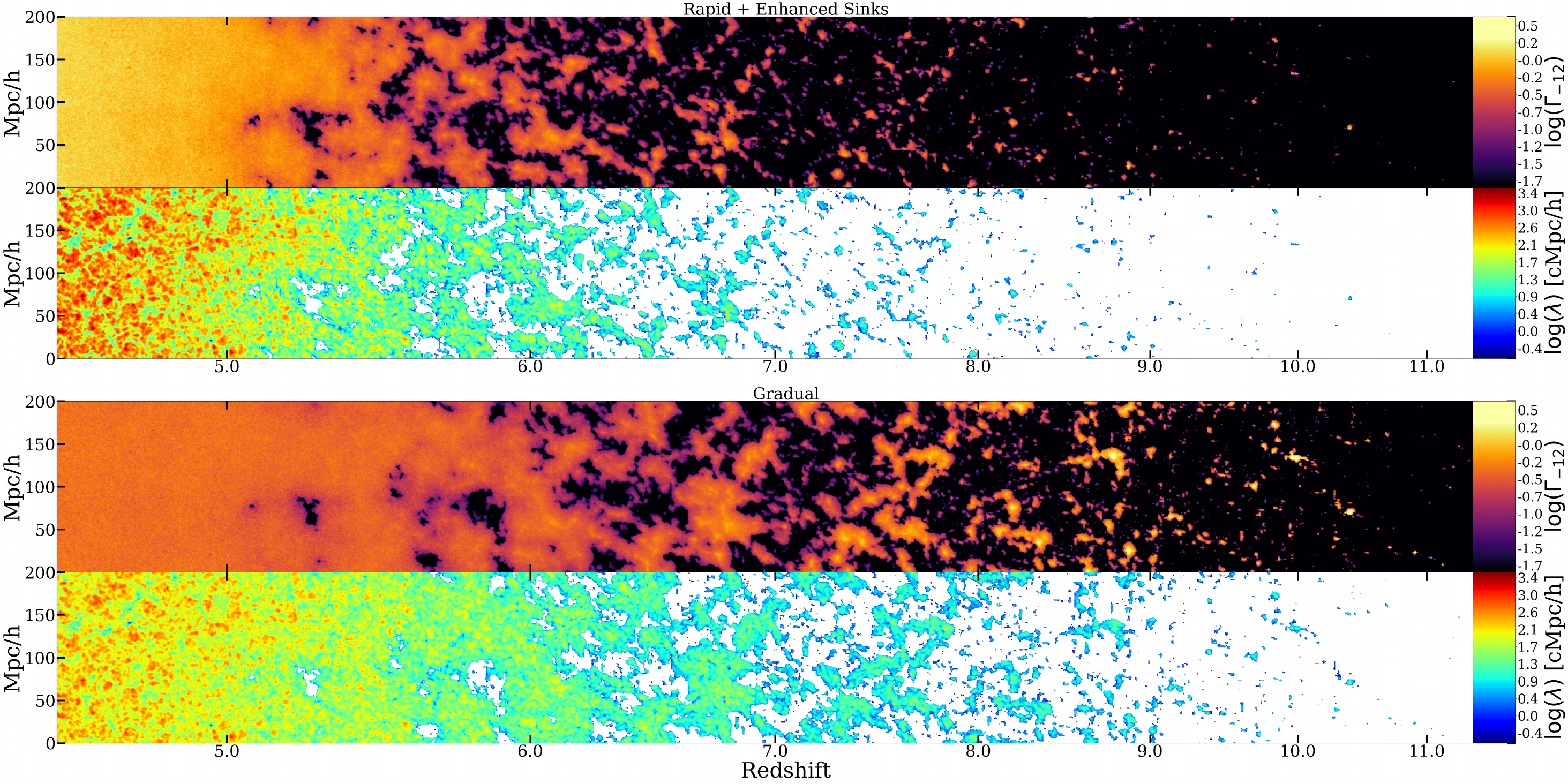}
    \caption{Light-cone slices (1 Mpc/h thick) of $\Gamma_{-12}$ and $\lambda_{912}^{\rm mfp}$ from our rapid+enhanced sinks (top) and gradual (bottom) models.  In the rapid model, $\lambda_{912}^{\rm mfp}$ evolves quickly at $z < 6$ as the gas relaxes and neutral islands disappear.  After reionization is finished, $\lambda_{912}^{\rm mfp}$ is limited by sinks in over-dense regions. In the gradual model most gas is relaxed by $z \sim 6$ and there is less neutral gas, resulting in a longer $\lambda_{912}^{\rm mfp}$.  }
    \label{fig:lightcone}
\end{figure*}

All models in Fig. \ref{fig:ion_history} end reionization at $z=5.1$ and formally have the same duration of $780$ Myr ($x_{\rm ion} = 1 - 99\%$).  The key distinction between the rapid and gradual scenarios is how quickly $x_{\rm ion}$ evolves in the last half of reionization.  In the rapid model (black/dotted), the emissivity peaks at $z \approx 6.5$ and reionization proceeds rapidly after its  $z=7.1$ midpoint.  As a result, $\Gamma_{-12}$ and $\lambda_{912}^{\rm mfp}$ grow rapidly between $z=6$ and 5.2, consistent with the measurements at those redshifts.   The cyan/dashed curve shows a model with the same emissivity, but with $\dot{n_{\gamma}} \propto L_{\rm UV}$.  In this case reionization is driven by rarer, brighter galaxies\footnote{For the $\dot{n}_{\gamma} \propto L_{\rm UV}$ model, half the ionizing photons are produced by halos with $M<M_{1/2}=(0.39,1.8,3.4) \times 10^{10}h^{-1}\mathrm{M}_{\odot}$ at $z=(9,6,5)$.  For the model with $\dot{n}_{\gamma}$ independent of $L_{\rm UV}$, production is peaked around $M_{\rm min}$; we find $M_{1/2}=(1.6,1.8,1.8) \times 10^{9}h^{-1}\mathrm{M}_{\odot}$.} which, on average, produce larger $\Gamma_{-12}$ in ionized bubbles.  This results in $\lambda_{912}^{\rm mfp}$ being too long across all redshifts. 

A principal conclusion from our modeling is that the short $\lambda_{912}^{\rm mfp}(z=6)$ measured by B21 prefers faint, less-biased sources as the main drivers of reionization. We have run a set of models with a lower $M_{\rm min}$ of $10^{8}h^{-1}M_{\odot}$.  Adopting the same $\dot{N}_{\gamma}$ as in our rapid model, we find $\Gamma_{-12} = 0.2$ and $\lambda_{912}^{\rm mfp}=5 h^{-1}$Mpc at $z=6$.   The dependence of $\lambda_{912}^{\rm mfp}$ on $M_{\rm min}$ can be understood in terms of halo bias.  Sources are less clustered in models with lower $M_{\rm min}$, which leads to ionized bubbles being smaller, on average.  Large-scale fluctuations in $\Gamma_{\rm HI}$ are also reduced. In contrast, for models with highly clustered sources, the intense ionizing radiation in over-dense regions quickly clears away the local sinks, allowing this radiation to penetrate much further into the IGM bulk.        

 In the gradual model (blue/dot-dashed), reionization proceeds more gradually after its $z = 8.5$ midpoint. By $z = 6$, more of the gas has relaxed in response to photoheating.  This, combined with the larger $x_{\rm ion}$, results in a factor of $1.5$ longer $\lambda_{912}^{\rm mfp}$ at $z=6$ compared to the rapid model. The evolution in $\Gamma_{-12}$ is flatter but it undershoots by a factor of $\sim 2$ the measurements of \citet{Becker2013} at $z<5$. Generally, models in which most of the IGM was reionized well before $z=6$ are difficult to reconcile with a short $\lambda_{912}^{\rm mfp}(z=6)$.  We have confirmed this with other runs as well, including those with $M_{\rm min}=10^8 h^{-1}$M$_\odot$.  The problem is that the local $\lambda_{912}^{\rm mfp}$ grow with time after $z_{\rm re}$ owing to relaxation and photoevaporation. In gas that was reionized at higher redshift, the only way to obtain low $\lambda_{912}^{\rm mfp}$ is to lower $\Gamma_{-12}$. But the slow evolution results in undershooting the Ly$\alpha$ forest measurements of $\Gamma_{-12}$ at $z\sim5$.  Hence, another principal conclusion from our modeling is that a rapidly evolving $x_{\rm ion}$ is required to recover both the short value of $\lambda_{912}^{\rm mfp}$ at $z=6$ and its rapid evolution to $z=5.2$. However, we note that the rapid models are in 2-3$\sigma$ tension with the \citet{McGreer2015} constraints on $x_{\rm ion}$ (top-left of Fig. \ref{fig:ion_history}). Updating these constraints with more QSO sight lines will provide a critical test of our assertion.  

There are two more obvious deficits of the rapid model (black/dashed): (1) The quick growth in $\lambda_{912}^{\rm mfp}$ continues below $z=5.2$, which is incompatible with measurements; (2) Relatedly, to control the growth of $\Gamma_{-12}$ and $\lambda_{912}^{\rm mfp}$ at $z<5.5$, $\dot{N}_{\gamma}$ must fall by $> 40\%$ in the $240$ Myr between $z=6$ and 5, a rapid evolution in the galaxy population (see however \citet{Ocvirk2021}). We emphasize that $\dot{N}_{\gamma}(z)$ is an input to our simulations; the decline is not the result of any feedback prescription.  D20 found that their $\lambda_{912}^{\rm mfp}$ are converged in resolution at the 10 \% (factor of $\sim 2$) level in relaxed (un-relaxed gas), respectively. Moreover, up to $\Delta t \approx 10$ Myr after $z_{\rm re}$, D20 found $\lambda_{912}^{\rm mfp}$ similar to the unheated simulations of \citet{Emberson2013}. This argues against numerical convergence being the sole culprit. 

One plausible explanation for the behavior of $\dot{N}_{\gamma}$ is that our sub-grid model overestimates $\lambda_{912}^{\rm mfp}$ in over-dense cells, e.g. by inadequately sampling massive sinks near the end of reionization.  To illustrate that we can obtain milder evolution in $\dot{N}_{\gamma}$ through sinks, we crudely scale down $\lambda_{912}^{\rm mfp}$ in all over-dense cells after $z=6.5$ by a factor of $[(1+z)/(1+6.5)]^{3.5}$, such that $\lambda_{912}^{\rm mfp}$ is a factor of 2 shorter in those cells by $z=5.2$.   The result is the rapid+enhanced sinks model (red/solid), for which $\dot{N}_{\gamma}$ levels off after $z=6.5$.  In this case, the enhanced sinks regulate the growth of $\Gamma_{-12}$ and $\lambda_{912}^{\rm mfp}$ so they do not outpace the measurements below $z=5.2$. This obviates the need for a rapid decline in $\dot{N}_{\gamma}$, illustrating an approximate degeneracy between the emissivity and the sinks.\footnote{We have also run an enhanced sinks version of the gradual model. While it is in better agreement with the $z<5$ $\Gamma_{-12}$ measurements -- reconciling a major deficiency of the model -- we find $\lambda_{912}^{\rm mfp}(z=6) = 14 h^{-1}$ Mpc.  This is still much larger than the measurement, suggesting that an even larger boost to the sinks at $z \sim 6$ would be required in this scenario.}

The thick curves in the top panel of Fig. \ref{fig:budget} show the cumulative number of photons per hydrogen atom absorbed.  The thin curves show the cumulative recombinations.\footnote{The number of recombinations is given by the total number of absorptions minus the net number of ionizations.  } For reference, the vertical line corresponds to reionization's end. All our models require $\approx 3$ photons/H atom to complete reionization. This is a factor of $1.5-2$ more than in recent models of the ionizing emissivity.\footnote{We find $1.5$ and $1.8$ photons/H atom in the models of \citet{Robertson2015} and \citet{Finkelstein2019}, respectively.}  Although more photons are absorbed earlier in the gradual model, the cumulative number is similar to the rapid models because much of the gas remains un-relaxed in the latter.  

 In a paper submitted concurrently with this work, \citet{Davies2021} quantify in detail the implications of the B21 measurement of $\lambda_{912}^{\rm mfp}(z=6)$ for high-$z$ galaxies. Considering also the dark pixel fraction constraints on $x_{\rm ion}(z=5.9)$, they find that $6.1^{+11.1}_{-2.4}$ photons per baryon are required to bring reionization to $90\%$ completion. Although this appears considerably larger than our budget, we note that $\lambda_{912}^{\rm mfp}(z=6)$ in our rapid models are $1\sigma$ longer than the central value of B21, and the neutral fractions are $x_{\rm HI} = 20\%$.  Adjusting for the former would bring down their budget to 3.7 photons per baryon.  Adjusting for latter would bring us further into agreement. Moreover, we have rerun the rapid simulation, but with a uniform $3.8\times$ ($2.5\times$) boost to the clumping (emissivity).  This yields $x_{\rm ion}(z=5.9) = 88 \%$ and $\lambda_{912}^{\rm mfp} =  3.65 h^{-1}$, closer to the B21 central value of $3.57 h^{-1}$ Mpc.  It requires 5.2 photons/H atom by $z = 5.9$.

The emissivity is commonly modeled as $\dot{N}_{\gamma} = f^{\rm eff}_{\rm esc} \xi_{\rm ion} \rho_{\rm UV}$, where $f^{\rm eff}_{\rm esc}$ is an effective escape fraction, $\xi_{\rm ion}$ is the ionizing efficiency, and $\rho_{\rm UV}$ is the UV luminosity density. The bottom panel of Fig.\ref{fig:budget} shows the product, $f^{\rm eff}_{\rm esc} \xi_{\rm ion}$, obtained by applying this relation to our $\dot{N}_{\gamma}$ and integrating the UV luminosity function of \citet{Finkelstein2019} for $\rho_{\rm UV}$. Previous studies have assumed $\log(\xi_{\rm ion}/[\mathrm{erg}^{-1} \mathrm{Hz}]) = 25.2-25.3$, consistent with the constraints of \citet{Bouwens2016} for $4 < z < 5$ galaxies.  There is evidence that the bluest galaxies at higher redshift exhibit higher efficiencies, $\log(\xi_{\rm ion}/[\mathrm{erg}^{-1} \mathrm{Hz}]) = 25.6-25.9$ \citep{Bouwens2016, Stark2017, Endsley2021}.  For the rapid models to be consistent with values of $25.8 (25.3)$ requires $f^{\rm eff}_{\rm esc}= 11 (35) \%$.  The gradual model requires a more extreme $f^{\rm eff}_{\rm esc}= 28 (89) \%$ at its peak of $z=10$.  However, if we let $M_{\rm min} = 10^{8} h^{-1}$M$_{\odot}$, we find $f^{\rm eff}_{\rm esc}= 10 (32) \%$, indicating that such an early peak in emissivity would likely require efficient star formation in galaxies with $M<10^9 h^{-1}$M$_{\odot}$ \citep[see e.g.][]{Finkelstein2019}.  Our $f^{\rm eff}_{\rm esc}$ are similar to those reported by~\citet[][]{Davies2021}. That our rapid models require $f^{\rm eff}_{\rm esc} = 10 - 40 \%$ supports the conclusion that faint $z>6$ galaxies must have been prolific leakers of LyC radiation, if reionization was driven by stellar emissions.    

The top and bottom sets of panels in Fig. \ref{fig:lightcone} show light cones through our rapid+enhanced sinks and gradual models, respectively.  In each set, the top(bottom) panel shows $\Gamma_{-12}$($\lambda_{912}^{\rm mfp}$).   The neutral islands down to $z\approx 5.5$ likely make both models compatible with the large opacity fluctuations observed in the Ly$\alpha$ forest \citep{Kulkarni2019, Nasir2020}. The Ly$\alpha$ forest mean flux evolution, however, may already disfavor the gradual model (c.f. bottom-middle of Fig. \ref{fig:ion_history}).   Another contrasting feature is the existence of large ($R \sim 10$s Mpc) ionized bubbles out to $z\sim 9$ in the gradual model, which is of interest for recent observations of bright LAE over-densities at $z>7$ \citep{Castellano2018, Tilvi2020, Endsley2021}. The different global $x_{\rm ion}$ and morphologies may be testable by forthcoming 21cm surveys. 

\section{Conclusion}
\label{sec:conc}

We have explored the implications of the B21 MFP measurements for reionization. Taken together with constraints on the intensity of the metagalactic ionizing background, we have argued that the rapid evolution from $\lambda^{912}_{\rm mfp}(z=6)=3.57^{+3.09}_{-2.14}$ cMpc/h to $\lambda^{912}_{\rm mfp}(z=5.1)=37.71^{+6.72}_{-5.06}$ cMpc/h favors a rapid and late reionization process. We have also argued that the short value of $\lambda^{912}_{\rm mfp}(z=6)$ is evidence that reionization was driven primarily by the faintest, least biased galaxies among its sources. In our preferred models, $\approx 3$ ionizing photons/H atom are required to complete reionization.  Half of them come from galaxies with $M\sim 10^{9}h^{-1}$ M$_{\rm \odot}$, or lower. At $z=6(8)$, this corresponds to UV magnitudes $M_{1600} \gtrsim -12.9 (-14.0)$. In addition to confirming the low value of $\lambda^{912}_{\rm mfp}(z=6)$, other avenues forward include updating the Ly$\alpha$ forest dark pixel limits on $x_{\rm ion}$ and constraining the IGM temperature at $z\sim 5.5$.  Our analysis highlights the complementary channels for constraining reionization with QSO absorption spectra.  

\acknowledgements

The authors thank Matt McQuinn, Fred Davies, Steve Furlanetto, Sarah Bosman, and Fahad Nasir for helpful comments on this work, and Hy Trac for providing his hydrodynamics code. AD was supported by HST-AR15013.005-A, NASA 19-ATP19-0191, and NSF AST-2045600. GDB was supported by NSF AST-1751404.  Computations were performed with NSF XSEDE allocation TG-AST120066 and NASA HEC Program through the NAS Division at Ames Research Center.  

\clearpage

\bibliography{references}{}
\bibliographystyle{aasjournal}

\end{document}